\documentclass[12pt]{article}
\input epsf.sty
\def\ZZZ{{\hbox{ Z\kern-1.6mm Z}}}

\newcommand{\tl}{\wt\lambda}

\newcommand{\wt}{\widetilde}

\newcommand{\TT}{{\cal T}}

\newcommand{\be}{\begin{equation}}
\newcommand{\ee}{\end{equation}}
\newcommand{\ben}{\begin{eqnarray}\displaystyle}
\newcommand{\een}{\end{eqnarray}}
\newcommand{\refb}[1]{(\ref{#1})}
\newcommand{\p}{\partial}

\def\one{{\hbox{ 1\kern-.8mm l}}}
\def\zero{{\hbox{ 0\kern-1.5mm 0}}}
 
\begin{document}
{}~
{}~
\hfill\vbox{\hbox{hep-th/0306137}
}\break
 
\vskip .6cm
\centerline{\Large \bf 
Open-Closed Duality at Tree Level}

\vskip .6cm
\medskip

\vspace*{4.0ex}
 
\centerline{\large \rm
Ashoke Sen}
 
\vspace*{4.0ex}

\centerline{\large \it Harish-Chandra Research Institute}

\centerline{\large \it  Chhatnag Road, Jhusi,
Allahabad 211019, INDIA}
 
\centerline{E-mail: ashoke.sen@cern.ch,
sen@mri.ernet.in}
 
\vspace*{5.0ex}
 
\centerline{\bf Abstract} \bigskip

We study decay of unstable D-branes in string theory in the presence of
electric field, and show that the classical open string theory results for
various properties of the final state agree with the properties of closed
string states into which the system is expected to decay. This suggests a
duality between tree level open string theory on unstable D-branes and
closed strings at high density.

\vfill \eject
 
\baselineskip=18pt

Recent studies involving decay of unstable D-branes in string theory show 
that at open string tree level such systems evolve to a configuration of 
zero pressure but non-zero energy density localized in the plane of the 
brane\cite{rolling}. Study of closed string emission from this system 
indicates
that 
once this process is taken into account the system decays into a set of 
highly non-relativistic, very massive closed 
strings\cite{0303139,0304192}. If we regard each of these closed strings 
as a {\it point-like} massive object without any internal structure, such 
a 
system of closed strings will have small ratio of pressure to 
energy 
density, and furthermore since the strings move very slowly away from the 
plane of the brane, the stress tensor of this system of closed 
strings will be localized close to the plane of the brane. This agrees 
with what is seen in the tree level open string analysis, and 
suggests 
that the tree level open string theory already knows about the fate of the 
closed strings that the system decays into\cite{0305011}.\footnote{
Of course, the precise final 
state will depend on the details of the quantum state of the initial 
D-brane\cite{0305011,0305159}, but this is presumably reflected in an 
ambiguity in the quantum corrections to the tree level open string 
results.}
There is, however, an intrinsic spread of the energy density away 
from the plane of the brane due 
to the internal oscillation of the final state closed strings. This effect 
does not seem to 
be captured 
by the tree level open string results. This may not necessarily be a 
contradiction, since defining the detailed profile of the 
energy-momentum tensor $T_{\mu\nu}$ 
requires us to make an ad hoc choice of the off-shell continuation of the 
graviton field. In contrast, the integrated $T_{\mu\nu}$ in the asymptotic 
future is the source of a (near)
on-shell zero momentum graviton\cite{9707068}, and for this there is 
perfect agreement 
between the open and the closed string results.

In this paper we study the decay of unstable D-branes in string 
theory in the presence of electric field. Such a system carries 
fundamental string winding charge along the direction of the electric 
field. The time evolution of the stress tensor and the winding charge 
density was determined in \cite{0208142,0301049} at open string 
tree 
level. 
In appropriate region in the parameter space we can estimate 
the stress tensor and winding charge density of the final state closed 
strings to which this system is expected to decay. We compare 
these with the asymptotic values of the answers obtained in tree level 
open string theory and again find 
agreement between the two sets of quantities. 
Although we shall not discuss this explicitly, the dilaton coupling 
computed from 
open and closed string descriptions
also agree, both being zero. This analysis 
therefore supports
the conjectured duality between the results in 
tree level open string theory on an unstable D-brane and the properties of 
the high density 
closed string 
states to which such D-branes decay. 
This is different from the usual open closed duality, in which the 
open string {\it loop amplitudes} contain information about the closed 
string poles\cite{loop}. 

Our analysis will be valid for D-branes in bosonic string theory, as well 
as 
non-BPS D-branes or brane-antibrane systems in type IIA or IIB 
string theory. 
We shall assume, for technical reasons, that the coordinates along 
the 
D-brane 
world-volume are compact so that the total energy and 
other charges of 
the 
system are finite. 
We shall denote by $x^1$ the coordinate along which the electric field 
on the non-BPS D-$p$-brane points,  by $x^i$ ($2\le i\le p$) the rest of 
the 
spatial world-volume coordinates on the D-$p$-brane, and by $\vec 
x_\perp$ the coordinates transverse to the D-$p$-brane. 
The classical solution describing the spatially homogeneous 
decay of the 
D-$p$-brane is parametrized by two parameters, 
the electric field $e$, and a constant $\tl$ labelling the initial 
position from which the tachyon starts rolling. In the 
asymptotic future
the 
components of the  
stress tensor $T^{\mu\nu}$ and the source $S^{\mu\nu}$ for the 
antisymmetric tensor field $B_{\mu\nu}$
approach the form\cite{0208142,0301049}
\ben \label{e4}
&& T^{00}=\Pi\, e^{-1} \, \delta(\vec x_\perp)\, , \quad
T^{11} = - \Pi \, e\, \delta(\vec x_\perp) \, , \quad
S^{01} = \Pi \, \delta(\vec x_\perp)\, , \nonumber \\
&&
T^{\mu\nu} =
S^{\mu\nu} =0 \quad \hbox{for other $\mu$, $\nu$}\, ,
\een
where
\be \label{e3}
\Pi\equiv e \,  (1-e^2)^{-1/2} \wt\TT_p \, \cos^2(\tl\pi) \, .
\ee
$\wt\TT_p$ denotes the (sum of the) tension(s) of the original D-branes.
In the limit
\be \label{elimit}
e\to 1, \qquad \tl\to{1\over 2}, \qquad \Pi \quad \hbox{fixed}\, ,
\ee 
we get\cite{0208142}
\be \label{e5}
T^{00}=\Pi\, \delta(\vec x_\perp)\, , \qquad
T^{11} = - \Pi \, \delta(\vec x_\perp) \, , \qquad
S^{01} = \Pi \, \delta(\vec x_\perp)\, .
\ee 
Since $|T^{00}|=|S^{01}|$, we see that 
the total energy available to the 
system is saturated by the BPS bound from the closed string winding 
charge.
This in turn implies that 
the only possible final closed string states into 
which the system can decay in this limit is a set of zero velocity closed 
strings wound along the direction $x^1$ without any oscillator 
excitations.
The stress-tensor and winding charge density associated with 
such 
a system of closed strings agree precisely with \refb{e5}\cite{0208142}. 
Thus we see 
that in this case the classical open string theory results \refb{e5} 
reproduce the 
expected property of the system of closed strings to which the system 
decays. In fact, the identification of these solutions with closed 
strings was 
first proposed in the context of effective field theory\cite{effective} 
before exact classical solutions were found using the techniques 
of boundary conformal field theory.

In what follows, we shall carry out a similar analysis without taking the 
limit \refb{elimit}. 
Since $\wt\TT_p\sim g^{-1}$, $g$ being the closed string coupling 
constant, we see from \refb{e3} that $\Pi$ is of order $g^{-1}$ for 
generic 
$e$ and $\tl$ of order one. We shall work in the weak coupling limit so 
that $|\Pi|$ is large, and assume, without any loss of generality, that 
$\Pi$ is positive.
Our goal will be to analyze
the final closed string state produced in the decay of this D-brane 
and check if the stress tensor and winding number density associated with 
this state agree with the tree level open string answers 
\refb{e4}. We note first of all that due to winding charge and energy 
conservation, the final state must carry a net winding charge and energy 
equal to that given by $S^{01}$ and $T^{00}$ in \refb{e4}. 
Thus the final state must have (a set of) closed strings wound along 
$x^1$. {}From \refb{e4} we also see that 
the 
system has an excess energy density $(T^{00}-S^{01})\propto 
\Pi(e^{-1}-1)$
beyond what is given by the BPS bound. 
In order to calculate properties of the final closed string state,
we need to understand 
where this excess energy density resides 
in the final state. We can 
consider two 
extreme cases:
\begin{enumerate}
\item All the excess energy density resides in the oscillation modes of 
`macroscopic' 
closed strings for which the (winding charge / energy) ratio is finite.
\item The decay product contains `microscopic' closed strings for which 
the winding charge is an insignificant fraction of their energy,
and
all the excess energy density resides in these microscopic closed string 
states.
\end{enumerate}
Of course we also have the more general possibility
where 
some part of the excess energy resides in the macroscopic strings and the
rest resides in the microscopic strings. We shall 
now present several (not totally independent)
arguments showing that for generic $e$ and $\tl$
possibility 1 is realized, {\it i.e.} the excees 
energy 
density almost fully resides in the macroscopic
strings. Furthermore in the
$g\to0$ limit
the transverse velocity of these strings vanish, and the (energy / 
winding charge) ratio of
each of these strings is given by
$e^{-1}$.
\begin{enumerate}
\item In \cite{0303139} it was shown that as the tachyon on a 
non-BPS D-brane rolls down towards the vacuum
in the absence of any electric field, the total 
energy carried by the closed strings produced during the decay is 
formally infinite. This in turn indicates that all the energy of the 
initial D-brane is carried away by these closed strings. In this analysis 
there was a delicate cancellation between the exponentially growing 
density of states of the final state closed strings and the exponentially 
suppressed amplitude for production of these strings. If we repeat the 
analysis for decay of a non-BPS D-brane with electric field, we get a 
finite answer (measured in $\alpha'=1$ unit) 
for the total energy carried by the microscopic closed strings produced 
during this process. This is simply a consequence of
the fact that the presence of the electric field slows down the decay 
process\cite{0208142} and 
hence gives rise to a larger exponential suppression for the production 
amplitude, whereas
the density of states remains the same. This indicates that 
the total fraction of energy of the initial 
D-brane 
that is carried away by the microscopic closed strings during the rolling 
of the tachyon is negligible in the weak coupling limit. Thus the energy 
must be somewhere else.

\item 
Consider a fundamental string wound (possibly multiple times) along $x^1$ 
carrying total winding charge $W$. If we excite such a string to 
oscillator level $N$, the energy of the state in $\alpha'=1$ unit 
is given by:
\be \label{e7}
E = \sqrt{W^2 + 4 N + \vec p_\perp^2}\, ,
\ee
where $\vec p_\perp$ is the transverse momentum. First we note 
that since the density of states grow exponentially with $\sqrt N$ but 
only as a power of $|\vec p_\perp|$, for a given energy 
it is `entropically favourable' to have $N>>|\vec p_\perp|^2$. In 
particular, by
following the argument  of \cite{0303139} one can show that we can 
maximize 
the density of states by taking $|\vec p_\perp|^2\sim \sqrt N$. This 
allows us to set $\vec p_\perp=0$ in \refb{e7} if we are willing to 
ignore terms of order unity in the expression for $E$. Then the density of 
closed string states grow as\footnote{The analysis of \cite{0303139} 
indicates that the actual number of closed string states available for the 
decay is only square root of the number given in \refb{e7a},
but the argument given below \refb{e7a} still holds.}
\be \label{e7a}
D(E,W)\sim \exp(2\beta_H\sqrt{N}) \sim \exp\left(\beta_H\sqrt{E^2 - 
W^2}\right)
\ee
where $\beta_H$ is the inverse Hagedorn temperature. In \refb{e7a} we 
have ignored multiplicative factor involving powers of $(E^2 - 
W^2)$.

Now we note that
if the same amount of energy and winding charge are distributed 
among two strings, -- one with energy $E_1$ and winding charge $W_1$, and 
the other with energy $E-E_1$ and winding charge $W-W_1$, then up to power 
law corrections the density 
of states of this system will be of order
\be \label{e7b}
D(E_1, W_1) D(E-E_1, W-W_1)\, ,
\ee
where $D(E,W)$ has been defined in \refb{e7a}. For fixed but large $W_1$, 
$E$ and 
$W$, this has a sharp
maximum at $E_1/W_1=E/W$.
Thus if the 
matrix elements for the decay 
of the D-brane into these states are comparable for different values of 
$E_1$, $W_1$, then the 
D-brane will decay predominantly into configurations of closed strings 
where the ratio of 
energy to the winding charge of each closed string is equal to the ratio 
of total energy 
to the total winding charge, {\it i.e.} $e^{-1}$. This, in particular,
shows that microscopic strings carry a 
negligible fraction 
of the final state energy.

\item Consider the decay of a static D0-brane. According 
to Ref.\cite{0303139} the final decay products are very massive closed 
strings 
with mass of order $g^{-1}$ and velocity of order $g^{1/2}$ where $g$ is 
the closed string coupling constant. Now consider boosting the system by a 
velocity $e$ in the $x^1$ direction. If the $x^1$ direction is 
non-compact, then due to the Lorentz invariance of the 
theory we would expect that in this case the final state closed strings 
will be moving with a velocity $e$ along the $x^1$ direction, with a small 
spread of order $g^{1/2}$ around $e$ due to the initial velocity of the 
decay products.
Now suppose the direction $x^1$ is compact so that the D0-brane is moving 
along a circle. Since boost along a circle direction is not a symmetry of 
the theory, we cannot strictly conclude that the final state closed 
strings will still be moving with a velocity $e$ along $x^1$, but let us 
assume that this is the case and proceed. In this case, the final 
state closed strings have energy/momentum ratio $e^{-1}$. Now making a 
T-duality 
transformation along $x^1$ converts the initial moving D0-brane into a 
D1-brane with electric field $e$ along it, and converts the final closed 
string states 
moving along $S^1$ into fundamental strings wound 
along 
$x^1$ with the energy/winding charge ratio $e^{-1}$.
Thus if we 
are allowed to use boost along a compact direction to study the system, we 
would conclude that most of the energy of the initial D1-brane is carried 
by final state closed strings which are wound along $x^1$, each with 
$T^{00}/S^{01}$ ratio $e^{-1}$.
This 
argument can be generalized to D-$p$-branes by making further T-duality 
transformation along the other directions.\footnote{Indeed, 
T-duality transformation and the assumption of boost 
invariance along a compact direction can be used to derive the 
results \refb{ecov4} from the corresponding results in the absence of 
winding charge.}

\end{enumerate}

Thus we arrive at the conclusion that the final state closed strings 
in the decay of a non-BPS D-brane with electric field along $x^1$ are 
predominantly closed strings wound along $x^1$, 
each carrying $T^{00}/S^{01}$ ratio $e^{-1}$.
We shall now compute the stress tensor and winding charge associated with 
such a system of closed strings and show that the result agrees with 
\refb{e4}.
For simplicity we shall carry out the 
calculation for the bosonic string, but it can be easily extended to 
include the fermionic terms in the action. 
We work in the covariant gauge where the degrees of freedom of the 
string are represented by $D$ free bosonic fields $X^\mu(\sigma,\tau)$, 
$D$ being the total dimension of space-time (26 for bosonic string 
theory). The stress tensor 
$T^{\mu\nu}$ and the winding charge density $S^{\mu\nu}$ are given by:
\ben \label{ecov1}
T^{\mu\nu}(y) &=& C \int d\sigma d\tau \, 
\prod_{\alpha=0}^{D-1}\delta(y^\alpha 
- X^\alpha(\sigma,\tau))\, (\p_\tau X^\mu \p_\tau X^\nu - \p_\sigma X^\mu 
\p_\sigma X^\nu) \, , \nonumber \\
S^{\mu\nu}(y) &=& C \int d\sigma d\tau \,
\prod_{\alpha=0}^{D-1}\delta(y^\alpha
- X^\alpha(\sigma,\tau))\, (\p_\tau X^\mu \p_\sigma X^\nu - \p_\tau X^\nu 
\p_\sigma X^\mu)\, ,
\een
for some constant $C$. 

The fields $X^\mu$ have expansion
\be \label{ecov2}
X^\mu = x^\mu + p^\mu \tau + w^\mu \sigma + X^\mu_{osc}\, , \qquad
X^\mu_{osc}\equiv \sum_{n\ne 0} {1\over n} \left(\alpha^\mu_n 
e^{in(\tau-\sigma)} + \bar \alpha^\mu_n
e^{in(\tau+\sigma)}\right)
\ee
where $p^\mu$ and $w^\mu$ denote respectively the momentum and winding 
along $x^\mu$, and $X^\mu_{osc}$ is the oscillator contribution. We shall 
focus 
on the case where the only non-zero components of $p^\mu$ and $w^\mu$ are 
$p^0$ and $w^1$ respectively, and also set the center of mass coordinates 
$x^\mu$ to zero.
We note first of all that due to the oscillation of the fundamental 
string in the transverse direction (the $X^i_{osc}$ part for $2\le i\le 
(D-1)$) the various charge densities clearly have spread in the 
transverse direction, and hence we shall not get strict localization of 
these charges on the plane of the brane. However, as argued earlier, 
this does not necessarily lead to a contradiction. Given this, we might as 
well look 
at the expression for the various sources after integrating them over the 
transverse directions $y^i$ for $2\le i\le (D-1)$. This removes the 
transverse 
$\delta$-functions from \refb{ecov1}. We can now use the remaining two 
delta functions to fix $\tau$ and $\sigma$, but we shall proceed in a 
different manner. Since we are interested in calculating the average 
property of the final state closed strings, we shall also integrate over 
the coordinates $y^0$ and $y^1$, and divide by the volume $V$ of 
space-time in 
the $y^0$-$y^1$ plane. This removes the remaining two delta functions and 
give:
\ben \label{ecov3}
\langle T^{\mu\nu} \rangle &=& C \, V^{-1} \, \int d\sigma d\tau \,
(\p_\tau X^\mu \p_\tau X^\nu - \p_\sigma X^\mu
\p_\sigma X^\nu) \, , \nonumber \\
\langle S^{\mu\nu} \rangle &=& C \, V^{-1} \, \int d\sigma d\tau \,
(\p_\tau X^\mu \p_\sigma X^\nu - \p_\tau X^\nu  
\p_\sigma X^\mu)\, ,
\een
where $\langle ~ \rangle$ denotes integration over the transverse 
directions and averaging over the $y^0$, $y^1$ directions.
We now note that the integral over $\sigma$ and $\tau$ make the 
contribution from the oscillator parts vanish. Thus we finally have:
\ben \label{ecov4}
&& \langle T^{00} \rangle = C \, V^{-1} (p^0)^2 \int d\sigma d\tau, \qquad
\langle T^{11} \rangle = -C \, V^{-1} (w^1)^2  \int d\sigma d\tau, 
\nonumber 
\\
&& \langle S^{01} \rangle = C \, V^{-1} p^0 w^1\int d\sigma d\tau, 
\nonumber 
\\
&& 
\langle T^{\mu\nu} \rangle =
\langle S^{\mu\nu}\rangle =0 \quad \hbox{for other $\mu$, $\nu$}\, .
\een
This form agrees with the classical 
open string 
theory result \refb{e4} provided we identify $w^1/ p^0$ as $e$. It is 
also clear that if we have a 
collection 
of such long strings (or multiply wound strings) with the same $w^1/
p^0$ ratio, then
the total 
stress tensor and winding charge density carried by the system, obtained 
by summing over \refb{ecov4} with different values of $p^0$ and 
$w^1=ep^0$, will 
also have the form given by \refb{e4}. 

This finishes our analysis showing that for the decay of D-branes in the 
presence of an electric field, the tree level open string theory 
reproduces the properties
of the closed strings produced in the decay of the 
D-brane. 
While the origin of this new duality is not 
completely 
clear, it could be related to the fact that there are no open string 
excitations around the tachyon vacuum. Thus if Ehrenfest theorem, -- which 
states that the classical results in a theory can be interpreted as 
quantum 
expectation 
values, -- has to work in this case, it must be that the tree level open 
string theory contain informtion about the properties of high 
density closed 
string states produced in the decay of the D-brane. This, in turn, 
suggests
that upon quantization this open string theory might contain closed 
strings in its spectrum, and could lead to a non-perturbative formulation 
of string theory. Recent observation\cite{0304224,0305159} that in the 
matrix 
model 
description of two dimensional 
string theory the matrix eigenvalue can be related to the open string 
tachyon on an unstable D0-brane provides strong evidence for 
this proposal.

{\bf Acknowledgement}: I would like to thank N.~Constable, R.~Gopakumar, 
J.~Karczmarek, Y.~Kim, J.~Minahan, S.~Minwalla, M.~Schnabl,
A.~Strominger, W.~Taylor, D.~Tong, N.~Toumbas, J.~Troost and 
B.~Zwiebach for useful discussion during the 
course of this work. I would also like to thank the Center 
for Theoretical Physics at MIT, where part of this work was done, for
hospitality and for providing a stimulating environment.

\end{document}